\documentclass[ floatfix,twocolumn]{revtex4}
\usepackage{amsmath}
\usepackage{color}
\usepackage{amssymb}

\usepackage{graphicx}

\begin{document}

\title{ Piezomagnetism in ferromagnetic superconductors}

\author{V.P.Mineev$^{1,2\footnote{E-mail: vladimir.mineev@cea.fr}}$}
\affiliation{$^1$Universite Grenoble Alpes, CEA, IRIG, PHELIQS, F-38000 Grenoble, France\\
$^2$Landau Institute for Theoretical Physics, 142432 Chernogolovka, Russia}

\begin{abstract}
The appearance of magnetization during the application of mechanical stress and the creation of elastic deformation during the application of a magnetic field are two fundamental properties of piezomagnetic materials. The symmetry of superconducting ferromagnets UGe$_2 $, URhGe, UCoGe allows piezomagnetism. In addition to conventional piezomagnetic properties occurring in both normal and superconducting states,
superconducting state of these piezomagnets has its own specificity.
Unlike conventional superconductors, in uranium ferromagnets, the critical transition temperature to the superconducting state changes its value when the direction of the field changes to the opposite.
\end{abstract}

\date{\today}
\maketitle
Piezomagnetism is an arising of magnetisation by applying to crystal a mechanical stress. Another manifestation of the same phenomenon is a linear magnetostriction - appearance of deformation linear in respect to magnetic field applied to the crystal \cite{LL}.
Piezomagnetism is possible in magnetic materials such that its group of symmetry ( magnetic class) does not contain the operation of time reversal or contains it  only in combination with rotations or reflections \cite{Tavger1956}. The first examples of such type substances was pointed out by I.E.Dzyaloshinskii \cite{Dzyal1957L}.
Soon after that the piezomagnetism was  observed 
in the antiferromagnetic fluorides of cobalt and manganese \cite{Borovik1960}. The more recent studies of this phenomenon and references   can be found in the work
\cite{Jaime2017}. 

The operation of time inversion in combination with rotations is not only found in magnetic classes of antiferromagnetic materials. This also takes place, for instance, 
in a ferromagnetic orthorhombic crystal with strong spin-orbital coupling fixing the
spontaneous magnetization along one of the symmetry axis of the second
order, say in the $z$-direction. Its point symmetry group 
 consists from the rotation on angle $\pi$ around $z$ axis
and the rotations on angle $\pi$ around $x$ and $y$ directions combined with the operation of time inversion $R$ which changes
the  direction of spontaneous magnetization to the opposite 
one, and also the operation of inversion $I$, the reflection $\sigma_h$ in the plane perpendicular to $z$ axis, and the reflections $\sigma_x,\sigma_y$ in the planes perpendicular to $x$ and $y$  axis combined with the operation
$R$
\begin{equation}
D_{2h}(D_{2})=
(E, C_{2}^{z}, RC_{2}^{x}, RC_{2}^{y},I,\sigma_h,R\sigma_x,R\sigma_y).
\label{e1}
\end{equation}
This is the point symmetry group of uranium ferromagnetic compounds UGe$_2$, URhGe and UCoGe intensively studied during the last two decades (see the recent reviews \cite{Flouquet2019,Mineev2017}). It should be clarified here that in the paper [7], the group 
$$
D_{2}(C_{2})=
(E, C_{2}^{z}, RC_{2}^{x}, RC_{2}^{y}),
$$
 which is a subgroup of the group $D_{2h}(D_{2})$, was incorrectly indicated as a point group of the ferromagnetic state of these substances. However, this does not lead to modifications of the theory of superconducting states developed in [7]. The fact is that the order parameters
superconducting states corresponding to irreducible and non-equivalent $A$ and $B$ co-representations of group $D_{2h}(D_{2})$ are given by the same formulas as for group $D_{2}(C_{2})$ (see equations (6) and (7) in the paper [7]).

The invariant in respect of transformations Eq.(\ref{e1}) thermodynamic potential   linear in respect to magnetic field and to the components of the stress tensor   is
\begin{equation}
\Phi=-\lambda_x\sigma_{xz}H_x-\lambda_y\sigma_{yz}H_y-(\lambda_{z1}\sigma_{xx}+\lambda_{z2}\sigma_{yy}+\lambda_{z3}\sigma_{zz})H_z.
\end{equation}
Thus, at application to the  crystal a mechanical stress along  of  each orthorhombic symmetry axis the crystal acquires  the additional magnetization
\begin{equation}
M_z=-\frac{\partial \Phi}{\partial H_z}=
(\lambda_{z1}\sigma_{xx}+\lambda_{z2}\sigma_{yy}+\lambda_{z3}\sigma_{zz}).
\end{equation}
As result, the fraction of bulk occupied by the ferromagnetic domains with magnetization parallel to the induced one increases.

The transversal components of magnetization appear at the application of twisting stress 
\begin{equation}
 M_x=-\frac{\partial \Phi}{\partial H_x}=\lambda_x\sigma_{xz},~~~~ M_y=-\frac{\partial \Phi}{\partial H_y}=\lambda_y\sigma_{yz}.
\end{equation}

An application of magnetic field along $z$-direction causes the appearance  the longitudinal deformations
\begin{eqnarray}
u_{xx}=-\frac{\partial \Phi}{\partial \sigma_{xx}}=\lambda_{z1}H_z,\\
u_{yy}=-\frac{\partial \Phi}{\partial \sigma_{yy}}=\lambda_{z2}H_z,\\u_{zz}=-\frac{\partial \Phi}{\partial \sigma_{zz}}=\lambda_{z3}H_z,
\end{eqnarray}
and along $x$ and $y$ directions the torsional deformations
\begin{equation}
u_{xz}=-\frac{\partial \Phi}{\partial \sigma_{xz}}=\lambda_xH_x,~~~~u_{yz}=-\frac{\partial \Phi}{\partial \sigma_{yz}}=\lambda_yH_y.
\end{equation}

The enumerated properties of  ferromagnetic UGe$_2$, URhGe and UCoGe are similar to properties of other piezomagnetic materials.
At temperature decrease these substances  pass to the superconducting state which  reveals specific
 piezomagnetic qualities. The superconducting state in these ferromagnetic metals is developed at least in two (spin-up, spin-down)  conducting bands 
 and described by multicomponent order parameter \cite{Mineev2017,Mineev2018,Mineev2020}. In order not to overload the presentation with the corresponding cumbersome calculations, we will demonstrate the manifestation of piezomagnetism in the superconducting state using the simplest one-component model of superconductivity.

 Let us consider the phase transition to the superconducting state in magnetic field  ${\bf H}$ lying in the $(x,z)$ plane.
 In high enough external magnetic field one can neglect by the acting on electron charges  internal field  generated 
 by the ferromagnetic magnetization and consider the specimen as the single domain specimen.
 The free energy functional quadratic in the superconducting order parameter  $\eta$ is 
 \begin{eqnarray}
 F=\int dV \left\{\alpha|\eta|^2\right.~~~~~~~~~~~~~~~~~~~~~~~\nonumber\\
 \left.+\tilde K_xD_x\eta(D_x\eta)^\star+\tilde K_yD_y\eta(D_y\eta)^\star+\tilde K_zD_z\eta(D_z\eta)^\star\right.\nonumber\\
 \left.
 +\frac{1}{2}\lambda H_x\left [D_x\eta(D_z\eta)^\star+D_z\eta(D_x\eta)^\star\right ] \right\},~~~~~~~~~
 \label{F}
 \end{eqnarray}
 Here, 
 \begin{eqnarray}
 \tilde K_i=K_i+\lambda_iH_z,~~~~i=x,y,z.
 \end{eqnarray}
 The  components of vector ${\bf D}=-i\nabla+\frac {2\pi}{\phi_0}{\bf A}$ are the operators of the long derivatives, $\phi_0=\pi\hbar c/e$ is the magnetic flux quantum,
  the vector-potential components are $A_x=-H_zy,~A_y=0,~A_z=H_xy$. All the terms of this functional  
 are  invariant in respect to all the transformations given by Eq.(\ref{e1}).
  
 The coefficient $\alpha=\alpha_0(T-T_{c0})$, where $T_{c0}$ is the "critical temperature" determined by the pairing interaction.  
   The pairing interaction in uranium ferromagnets depends on magnetic field, hence $T_{c0}$ is also field dependent quantity \cite{Mineev2017}.
   Depending on the field magnitude and its direction the equilibrium value of "critical temperature" is not changed if an applied magnetic field is reversed in direction 
   \begin{equation}
   T_{c0}({\bf H})=T_{c0}(-{\bf H}).
   \end{equation}
   The real critical temperature of transition to superconducting state in magnetic field  is determined by the field dependence $T_{c0}({\bf H})$ and the orbital effects 
   to which we proceed.

  The corresponding to the functional (\ref{F}) Ginzburg-Landau equation is
  \begin{eqnarray}
 \alpha\eta
 -\tilde K_x\left (\frac{\partial}{\partial x}-\frac{2\pi i}{\phi_0}H_zy\right )^2\eta\nonumber\\
 -\tilde K_y\frac{\partial^2\eta}{\partial y^2}
 -\tilde K_z\left (\frac{\partial}{\partial z}+\frac{2\pi i}{\phi_0}H_xy\right )^2\eta\nonumber\\-\lambda H_x\left (\frac{\partial}{\partial x}-\frac{2\pi i}{\phi_0}H_zy\right )
 \left (\frac{\partial}{\partial z}+\frac{2\pi i}{\phi_0}H_xy\right )\eta=0.
 \label{GL}
 \end{eqnarray}
The solution of this equation has the form
\begin{equation}
\eta(x,y,z)=\exp({iq_xx}+{iq_zz})\psi(y)
\end{equation}
 For the upper critical field determination it is enough to find the solution 
 at $q_x=q_z=0$. In this case, the equation  for the function $\psi(y)$ is
 \begin{eqnarray}
 \alpha\psi+\left (\frac{2\pi F}{\phi_0\tilde K_y^{1/2}}\right)^2y^2\psi
  -\tilde K_y\frac{\partial^2\psi}{\partial y^2}=0,
 \label{u}
 \end{eqnarray}
 where 
 \begin{equation}
 F=F({\bf H})=\left \{\tilde K_y\left [\tilde K_xH^2_z+\tilde K_zH^2_x-\lambda H_zH_x^2\right ]\right \}^{1/2}.
 \end{equation}
 The solution is
 \begin{equation}
 \psi(y)=\exp\left (- \frac{\pi F}{\phi_0\tilde K_y}y^2  \right   )
 \end{equation}
 if the following relation is met
\begin{equation}
F({\bf H})=-\frac{\phi_0}{2\pi}\alpha,
\end{equation}
 which can be also rewritten as the field dependence of the critical transition temperature to the superconducting state
   \begin{equation}
T_c({\bf H}) =T_{c0}({\bf H})-\frac{2\pi}{\phi_0\alpha_0}\left \{\tilde K_y\left [\tilde K_xH^2_z+\tilde K_zH^2_x-\lambda H_zH_x^2\right ]\right \}^{1/2}.
\end{equation}
We see that except for the field direction parallel to $x$ axis the  critical temperature  changes if an applied magnetic field is reversed in direction 
\begin{equation}
T_{c}({\bf H})\ne T_{c}(-{\bf H}).
\end{equation}

In summary, it is shown that in ferromagnetic crystals UGe$_ 2$, URhGe and UCoGe in an external magnetic field arise
the deformations (5)-(8) directly proportional to the
field, and in the absence of the field the mechanical stresses (3),(4) cause the magnetization. In contrast to conventional superconductors, in the uranium ferromagnets the transition temperature in superconducting  state in a magnetic field (18) changes when the direction of the external field is reversed.

\end{document}